\documentclass[aps,prl,twocolumn,superscriptaddress,showpacs]{revtex4-1}
\usepackage{amsmath,amssymb,bm,graphicx,url,epsf,color}
\usepackage{wasysym}
\usepackage{MnSymbol}

\begin{document}

\title{Chargeless heat transport in the fractional quantum Hall regime}

\author{C. Altimiras}
\thanks{Current address: CEA, Service de Physique de l'\'Etat Condens\'e (SPEC), 91191 Gif-sur-Yvette, France}
\affiliation{CNRS / Univ Paris Diderot (Sorbonne Paris Cit\'e), Laboratoire de Photonique et de Nanostructures
(LPN), route de Nozay, 91460 Marcoussis, France}
\author{H. le Sueur}
\thanks{Current address: CNRS, Centre de Spectrom\'etrie Nucléaire et de Spectrom\'etrie de Masse (CSNSM), 91405 Orsay Campus, France}
\affiliation{CNRS / Univ Paris Diderot (Sorbonne Paris Cit\'e), Laboratoire de Photonique et de Nanostructures
(LPN), route de Nozay, 91460 Marcoussis, France}
\author{U. Gennser}
\affiliation{CNRS / Univ Paris Diderot (Sorbonne Paris Cit\'e), Laboratoire de Photonique et de Nanostructures
(LPN), route de Nozay, 91460 Marcoussis, France}
\author{A. Anthore}
\affiliation{CNRS / Univ Paris Diderot (Sorbonne Paris Cit\'e), Laboratoire de Photonique et de Nanostructures
(LPN), route de Nozay, 91460 Marcoussis, France}
\author{A. Cavanna}
\affiliation{CNRS / Univ Paris Diderot (Sorbonne Paris Cit\'e), Laboratoire de Photonique et de Nanostructures
(LPN), route de Nozay, 91460 Marcoussis, France}
\author{D. Mailly}
\affiliation{CNRS / Univ Paris Diderot (Sorbonne Paris Cit\'e), Laboratoire de Photonique et de Nanostructures
(LPN), route de Nozay, 91460 Marcoussis, France}
\author{F. Pierre}
\email[Corresponding author: ]{frederic.pierre@lpn.cnrs.fr}
\affiliation{CNRS / Univ Paris Diderot (Sorbonne Paris Cit\'e), Laboratoire de Photonique et de Nanostructures
(LPN), route de Nozay, 91460 Marcoussis, France}

\date{\today}

\begin{abstract}
We demonstrate a direct approach to investigate heat transport in the fractional quantum Hall regime. At filling factor of $\nu=4/3$, we inject power at quantum point contacts and detect the related heating from the activated current through a quantum dot. The experiment reveals a chargeless heat transport from a significant heating that occurs upstream of the power injection point, in the absence of a concomitant electrical current. By tuning \emph{in-situ} the edge path, we show that the chargeless heat transport does not follow the reverse direction of the electrical current path along the edge. This unexpected heat conduction, whose mechanism remains to be elucidated, may play an important role in the physics of the fractional quantum Hall regime.
\end{abstract}


\maketitle

The quantum Hall effect arises for two-dimensional electrons subjected to a strong perpendicular magnetic field and involves gapless electronic excitations propagating in channels along the sample edge \cite{wen1992tesfqh}. It is evidenced from distinct plateaus in the Hall resistance $R_H=R_K/\nu$, with $R_K=h/e^2$ the resistance quantum, accompanied by a vanishing longitudinal resistance. At fractional values of the filling factor $\nu$, this effect is due to Coulomb interaction. It is associated with the formation of exotic electronic phases \cite{laughlin1983aqhe}, with quasiparticle excitations markedly different from bosons and fermions and carrying a fraction of the electron charge \cite{saminadayar1997oes3fclq,depicciotto1997dofc}. Although the fractional quantum Hall effect was discovered three decades ago \cite{tsui1982fqhe}, the experimental investigation of many striking aspects of this physics is still at an incipient stage. This includes the predicted anyonic \cite{wen1992tesfqh} and possibly non-abelian statistics \cite{moore1991nafqhe} of the fractional quasiparticles, and the presence of correlated electronic edge modes carrying heat but no charge \cite{kane1994nu2s3,lee2007phs5s2,levin2007phspf5s2}.

\begin{figure}[!htbp]
\centering\includegraphics[width=1\columnwidth]{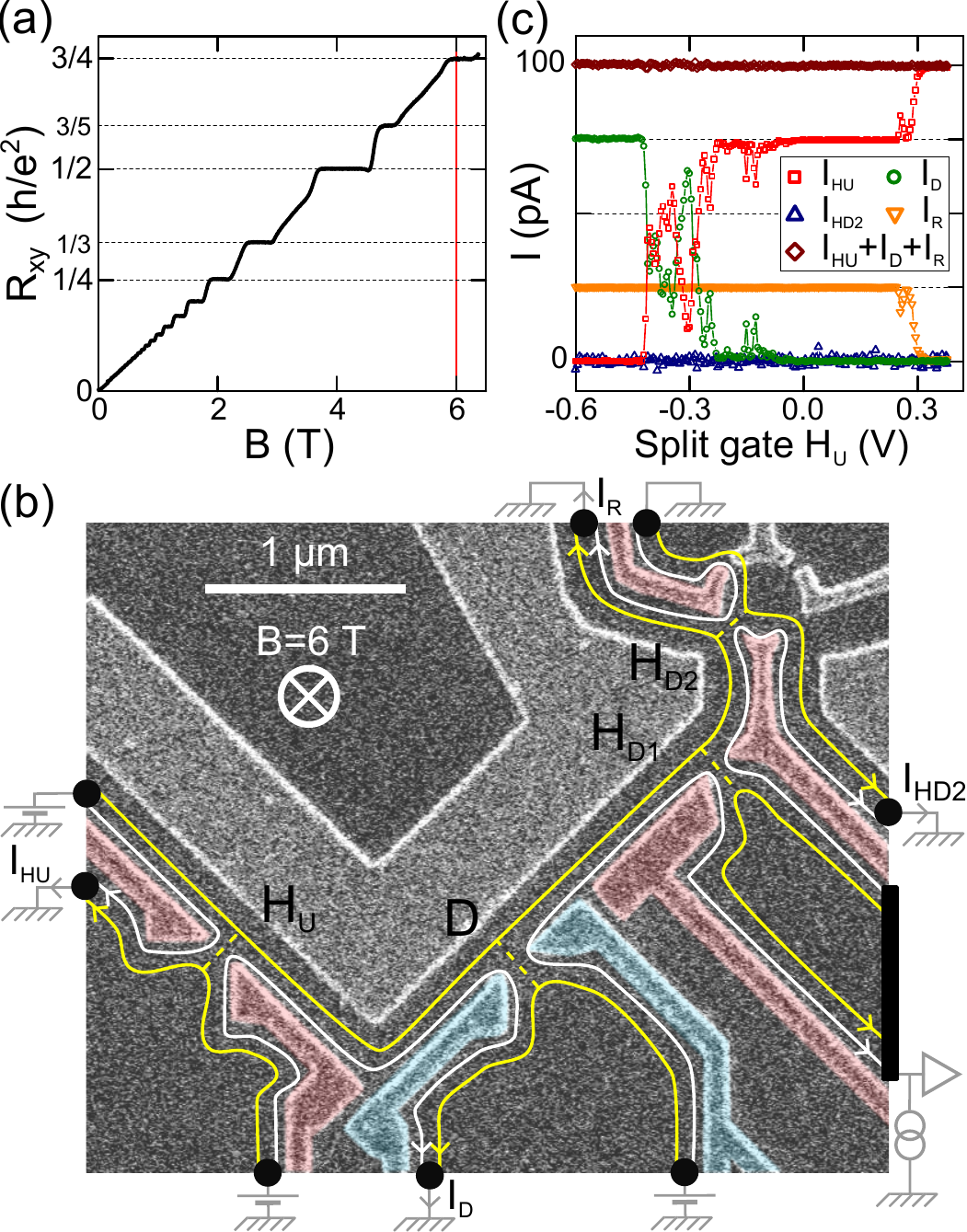}
\caption{
(color online). Measured device. (a) Measured Hall resistance $R_{xy}$ versus perpendicular magnetic field $B$. (b) SEM micrograph of the sample. Surface metal gates appear brighter. Large and top right gates (not colorized) are grounded. The large ohmic contacts indicated as black disks or as a rectangle are located hundreds of microns away. At filling factor $\nu=4/3$, the electrical current propagates anticlockwise along two edge channels (lines). The fractional inner channel (yellow line) is locally heated up by voltage biasing a quantum point contact (upstream $H_U$, downstream $H_{D1}$ or $H_{D2}$) using its split gate (colorized red) to set it to half (fully) transmit the inner (outer) channel. The heating induced in the inner channel is probed from the current $I_D$ across the detector ($D$) set in the Coulomb blockade regime using its split gate (colorized blue). (c) Test of electrical edge paths, with $H_{D2}$ and $D$ set to perfectly transmitting the integer outer channel, $H_{D1}$ set to perfectly reflecting it, and the fractional inner channel being fully reflected at the three split gates. Symbols display outgoing currents, measured at different locations versus the gate voltage controlling $H_U$, for a current of $100~\mathrm{pA}$ injected at the left top contact.
}
\label{fig1}
\end{figure}

It was pointed out since the mid-nineties that the study of heat transport would provide decisive information on the peculiar physics of the different fractional quantum Hall regimes \cite{kane1997qttfqhe,levin2007phspf5s2,grosfeld_probingneutral_2009,fertig2009vfe,takei2010neqesll,viola2012tpnemfqhr}. Very recently, a non-chiral heat transport at several fractional filling factors was evidenced using noise measurements, and attributed to the presence of upstream neutral edge modes \cite{bid2010obsnm,dolev2011cnmfssll,gross2011unmfquehwcd}. In the present work, we demonstrate a direct approach to investigate heat transport in the fractional quantum Hall regime at the filling factor $\nu=4/3$ (Fig.~1(a)). For this purpose we controllably inject power at several locations along the sample channel, using voltage biased quantum point contacts, and detect the resulting heating from the thermally activated current across a quantum dot located at an intermediate edge position (Fig.~1(b)). With this approach, we first evidence an unexpected heating upstream power injection, with respect to the chiral electrical current along the edge. We then demonstrate that this chargeless heat current flows in the bulk, further away from the edge than the electrical path. The relatively important upstream heating suggests the corresponding chargeless heat transport mechanism may play an important role in the physics of the fractional quantum Hall regime.

The studied sample is tailored in a typical two-dimensional electron gas of density $2~10^{15}~\mathrm{m}^{-2}$ and mobility $250~\mathrm{m}^2\mathrm{V}^{-1}\mathrm{s}^{-1}$, buried 105~nm deep in a GaAs/Ga(Al)As heterojunction. Note that similar observations on a second sample confirmed the reported findings. We performed the measurements either at DC or by standard lock-in techniques at frequencies below $100~$Hz, in a dilution refrigerator of base temperature 40~mK \cite{si}. Heaters, detector and sample geometry are tuned by field effect using capacitively coupled surface metal gates (Fig.~1(b)). We applied a perpendicular magnetic field $B=6.0~$T to set the sample in the middle of the zero longitudinal resistance plateau at $\nu=4/3$ (see Fig.~1(a) and \cite{si}, the extracted thermal activation transport gap is $\sim k_B \times 700$~mK). According to the effective edge state theory \cite{wen1992tesfqh}, the electrical edge current at this bulk filling factor is carried by two channels co-propagating in the same direction. The `$\nu=1$' outer channel (white line in Fig.~1(b)) is associated to the integer quantum Hall physics, and the `$\nu=1/3$' inner channel to the fractional physics (yellow line in Fig.~1(b)).

The data in Fig.~1(c) confirms the reality of the above edge picture. A bias of $1.9~\mu\textrm{V}\simeq(3/4)R_K\times100~$pA is applied to the left top contact and the resulting currents are measured at different locations as a function of the split gate voltage tuning the constriction $H_U$. The current $I_{HU}$ transmitted across $H_U$ is zero for gate voltages below $-0.5~$V and increases up to the injected current above $0.3~V$. Importantly, $I_{HU}$ shows a wide plateau, larger than $0.3~V$, at $3/4$ of the injected current. This plateau corresponds to the full transmission of the `$\nu=1$' outer channel, which carries three times more current than the fully reflected fractional `$\nu=1/3$' inner channel. Similar behaviors are observed across all the studied constrictions of this sample. In order to establish the distinctness of the two copropagating channels, the constrictions labeled $D$ and $H_{D2}$ were tuned to fully transmitting (reflecting) the outer (inner) channel, and the constriction $H_{D1}$ was closed. In this regime, the vanishing currents $I_D$ and $I_{HD2}$ at positive split gate voltages show that the electrical current $I_R$ is carried only by the inner channel, with negligible charge tunneling toward the outer channel between $H_U$ and $H_{D2}$. Similar tests were performed in presence of the largest injected powers to establish the counter-clockwise (chiral) propagation of the electrical current as well as the absence of inter-channel tunneling in all the experimental configurations investigated hereafter.

\begin{figure}[!htbp]
\centering\includegraphics[width=1\columnwidth]{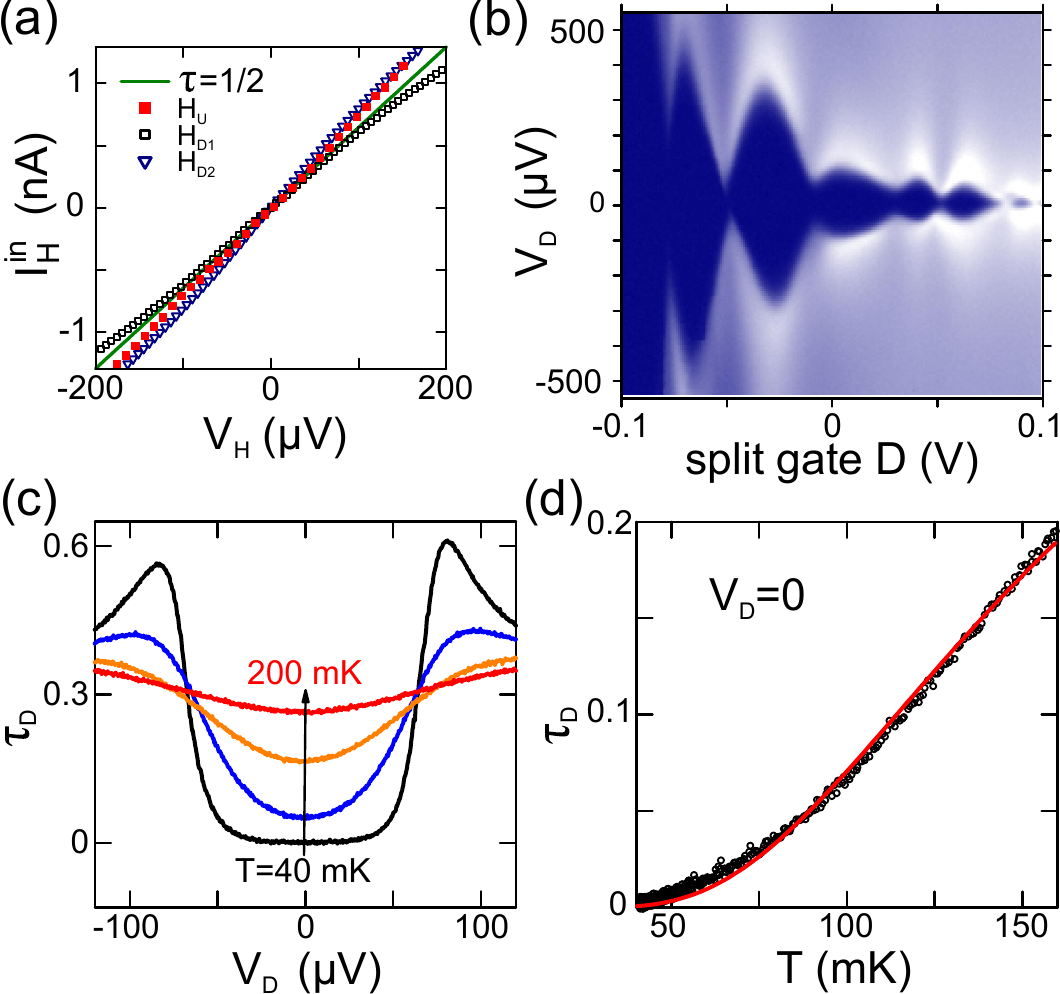}
\caption{(color online). Heaters and detector characterizations. (a) $I-V$ characteristics of the heater quantum point contacts (symbols), with $I^{in}_{H}$ the measured DC current transmitted in the inner edge channel and $V_{H}$ the applied voltage bias. The continuous line (green) is a calculation for an exactly half transmitted inner edge channel. (b)   Representative surface plot of $\partial I_{D}/\partial V_D\equiv \tau_D/3R_K+1/R_K$ (higher is brighter) versus the detector bias voltage $V_D$ and the split gate voltage controlling $D$. Darkest areas correspond to the detector inner edge channel transmission $\tau_D=0$. (c)  Measured $\tau_D$ versus $V_D$ for different temperatures $T\in \{40,100,150,200\}$~mK, keeping the detector split gate voltage fixed. (d) Symbols: measured $\tau_D(V_D=0)$ versus $T$ for the same detector settings as the data in (c). Continuous line (red): fit assuming a metallic quantum dot (see text).
}
\label{fig2}
\end{figure}

Now that we have characterized charge transport, we investigate heat transport by injecting power and probing the resulting heating in the fractional inner edge channel.

Power is injected locally into the inner channel by applying a voltage bias $V_H$ across a constriction set to transmit half (all) of the current carried by the inner (outer) channel. These constrictions were tuned to have little voltage dependence of their transmission, as shown Fig.~2(a). At half transmission of the inner channel, the injected power into edge excitations is $P_H\simeq0.25V_H^2/6R_K$ \cite{si}. One heater $H_{U}$ is located at an edge distance $1.8~\mu$m upstream the detector $D$, and two heaters $H_{D1}$ and $H_{D2}$ are located respectively at $1.4$ and $2.2~\mu$m downstream the edge channel.

Heating in the fractional inner channel is detected from the activated ac current across the constriction $D$ tuned to the Coulomb blockade regime. Interestingly, this regime is here obtained with a simple split gate. Such a behavior is usually attributed to small variations of the 2DEG density of states in the vicinity of the constriction. The Coulomb blockade regime is first evidenced by the appearance of Coulomb diamonds in the bias and gate voltage dependence of the inner channel transmission $\tau_D \equiv 3R_K\partial I_D / \partial V_D -3$ (Fig.~2(b)). The detector split gate voltage is adjusted using the Coulomb diamonds to tune the activation temperature of $\tau_D$ to a value much higher than the base temperature, but sufficiently low to detect a small heating. Figure~2(c) displays $\tau_D$ versus the detector voltage bias and for several temperatures, at the working point used hereafter (unless otherwise specified). In order to minimize power injection at the detector, we probe the electronic temperature from $\tau_D$ measured at zero detector bias voltage $V_D=0$ and we restrict our investigation to $\tau_D<0.2$. The detector is calibrated at thermal equilibrium by measuring $\tau_D$ versus temperature $T$ (Fig.~2(d)). Remarkably, despite the fractional character of the studied `$\nu=1/3$' inner edge channel, we find a very good agreement between the measured $\tau_D(T)$ (symbols) and the simple Coulomb thermometry expression \cite{beenakker1991tcbocqd,kouwenhoven1997etqd} $\tau_D \propto \cosh^{-2}(T_C/T)$ (red line) with an activation temperature $T_C=155~$mK, compatible with the non-linear characterization value $\simeq 150~$mK \cite{si}. In the present heat transport experiment, only one side of the detector is heated up and the electronic energy distribution in the corresponding inner edge channel could be different from an equilibrium distribution function \cite{altimiras2010nesiqhr}. We therefore extract an effective temperature $T_{eff}$ from the inverted temperature calibration shown Fig.~2(d).

\begin{figure}[!htbp]
\centering\includegraphics[width=1\columnwidth]{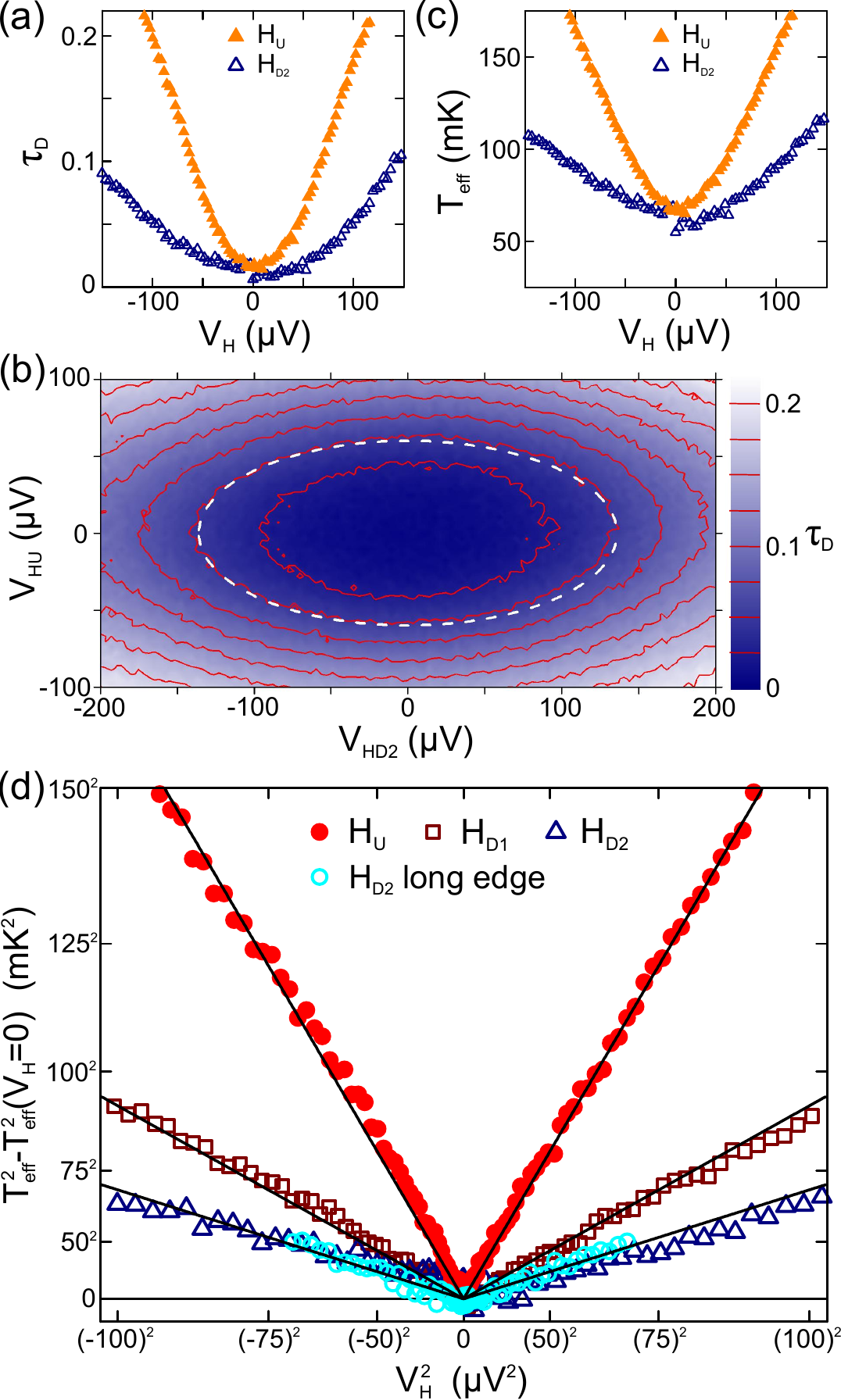}
\caption{
(color online). Heat detection versus heater position. (a) Measured detector inner edge channel transmission $\tau_D(V_D=0)$ versus the voltage bias $V_H$ applied either to the heater $H_U$ upstream ($\blacktriangle$) or $H_{D2}$ downstream ($\triangle$), with the inner edge channel fully reflected at $H_{D1}$. (b) Surface plot of $\tau_D$ versus the simultaneously applied upstream $V_{HU}$ and downstream $V_{HD2}$ heater voltages, with the inner channel reflected at $H_{D1}$. The detector is here set to a slightly higher activation energy than elsewhere. Continuous lines (red) are equitransmission contours at integer multiples of 0.025. The dashed line is a fit of the equitransmission line $\tau_D=0.05$  with an ellipse of minor (major) diameter $60~\mu$V ($136~\mu$V) along $V_{HU}$ ($V_{HD2}$). (c) Effective temperature $T_{eff}$ extracted from $\tau_D(V_D=0)$ in (a), using the  temperature detector calibration (Fig.~2(d)). (d) Symbols: Difference between the squared effective temperature and the thermal contribution $T_{eff}^2(V_H=0)$ ($\propto$ energy increase) plotted versus $V_H^2$ ($\propto$ injected power) for each heater. Full (open) symbols correspond to heating upstream (downstream). In the `$H_{D2}$ long edge' configuration, more than $91~\%$ of the inner channel electrical current is deviated toward a large ohmic contact at the intermediate downstream heater $H_{D1}$. Straight lines are guides to the eye.
\label{fig3}
}
\end{figure}

Figure 3(a) shows the detector transmission as a function of the voltage bias $V_H$ applied either to the heater upstream ($H_U$, $\blacktriangle$) or downstream ($H_{D2}$, $\vartriangle$), with the inner edge channel fully reflected at $H_{D1}$. Assuming only that $\tau_D$ increases with the temperature, we find as expected that heating is the largest for the upstream heater $H_U$, directly connected to the detector by the current carrying edge modes, and that heating increases with $V_H$. More surprisingly, the raw data demonstrates the presence of a smaller but relatively important heating from the downstream heater $H_{D2}$, without associated electrical current. This is in contrast with the co-propagation of heat and charge seen at integer filling factors \cite{granger2009och,bid2010obsnm}, and in particular on the same sample at $\nu=2$ \cite{altimiras2010nesiqhr,altimiras2010ctrlrelaxqhr}. Figure~3(b) shows on a surface plot of $\tau_D$ the interplay of heating upstream and downstream the detector (the detector is here set to a slightly higher activation temperature than elsewhere). Remarkably, the equitransmission lines (red) display ellipsoid shapes of similar aspect ratios. This is illustrated at $\tau_D=0.05$ with the ellipse $(V_{HU}/60~\mu \textrm{V})^2+(V_{HD2}/136~\mu \textrm{V})^2=1$ (white dashed line). The detected heating is therefore approximately given by simply summing up the upstream and downstream injected power with a fixed scaling factor. The observation that injecting power in the upstream heater does not facilitate the chargeless heating from $H_{D2}$ confirms that this phenomena is not related to a local destruction of the fractional state.

To investigate further the heat transport mechanisms, and in particular the chargeless heat transport possibly driven by neutral excitations, the effective temperature $T_{eff}$ is extracted from the measured $\tau_D$ (Fig.~3(c)). In order to focus on the increase in energy density within the `$\nu=1/3$' inner edge channel, we plot for different heater positions $T_{eff}^2(V_H)-T_{eff}^2(V_H=0)$ as a function of $V_H^2$, which is proportional to the injected power (Fig.~3(d)). The validity of this procedure to subtract the thermal background was established experimentally by checking that data taken at different temperatures $T=\{50,100,150\}~$mK fall on top of each other \cite{si}. Note that, if we assume a thermal energy distribution at the temperature $T_{eff}(V_H)$, the plotted quantity would be directly proportional to the increase in electronic energy density due to the injected power. The increase in the effective energy density is found proportional to the injected power when heating upstream ($H_U$), as expected in the simple edge channel picture \cite{altimiras2010nesiqhr}. Interestingly, the same linear dependence is also observed when heating downstream the detection point ($H_{D1}$, $H_{D2}$), in presence of only chargeless heat transport (straight lines are guides to the eye). These observations were reproduced for different settings of the heat detector and on two samples. They are compatible with a proportion of injected power transferred at the heaters into neutral modes, which does not depend on energy (nor on base temperature up to 150~mK \cite{si}). It is also consistent with a chargeless heat current that has the same energy dependence as the heat current by the charged edge modes, which is expected to be proportional to the energy density. Interestingly, the same heating is detected when using either the closest downstream heater $H_{D1}$, or the furthest downstream $H_{D2}$ with an injected power increased by a factor $1.8\pm0.3$, similar to the heater-detector distance ratio $\sim1.6$ (this quantitative comparison can be done using the raw $\tau_D$ directly).

Neutral edge modes propagating in the opposite direction to the electrical current are not usually expected at $\nu=4/3$ \cite{kane1995istfqhes}. Nonetheless, such phenomena could result from edge reconstruction due to Coulomb interaction in presence of a realistic smooth confinement potential at the edge \cite{macdonald1990esfqhe,aleiner1994nee,chamon1994er,wan2002rfqhe}. In order to discriminate between chargeless heat transport along the edge or through the bulk, we deviate the electrical edge path between the detector and the heater $H_{D2}$ toward a macroscopic ohmic contact located six hundred microns away. This is done by opening the intermediate constriction $H_{D1}$. Note that the same ohmic contact at $\nu=2$ was found to behave like a reservoir of cold electrons \cite{altimiras2010ctrlrelaxqhr}. Here the simultaneous monitoring of the conductance through $H_{D1}$ allows us to ascertain that between $91~\%$ and $96~\%$ of the electrical current carried by the inner edge channel reaches the contact. Therefore, if the chargeless heat transport is carried by neutral modes following the reverse direction of the electrical current along the edge, we should observe a strong reduction in the detected heating. On the contrary, the corresponding data labeled `$H_{D2}$ long edge' in Fig.~3(d) ($\circ$ ) are indistinguishable, at our relative experimental accuracy of $\pm15\%$ \cite{si}, from injecting power with the same heater $H_{D2}$ without deviating the edge path ($\vartriangle$). This shows that the presently observed chargeless heat transport propagates through the bulk. We remark that this central conclusion can be reached directly from the raw $\tau_D$ measurements. Note also that the observation of a similar upstream heat signal, when the injected power and heater-detector distance are scaled by the same factor, is consistent with an isotropic 2D-bulk heat transport (see \cite{si} for further discussions on heat paths). Intriguingly, the recent noise measurements investigating neutral edge modes \cite{bid2010obsnm,dolev2011cnmfssll,gross2011unmfquehwcd} have not pointed out such a chargeless heat transport through the bulk. However, to the best of our knowledge, these previous noise measurements would not discriminate between bulk and edge heat transport \cite{si}.

The mechanism responsible for the presently observed chargeless heat transport is presently not known. In principle the coupling to phonons is possible, but different estimates suggest it is negligible \cite{martin1990ephqhr,price1982eph2deg,si} and it did not result in discernable heat transfers at $\nu=2$ on the same sample and energy scales for propagation distances up to $30~\mu$m \cite{lesueur2010relaxiqhr,altimiras2010ctrlrelaxqhr,si}. Heat transfers between edge states and the electronic excitations in the nearby surface metallic gates were also found negligible at $\nu=2$ \cite{lesueur2010relaxiqhr,altimiras2010ctrlrelaxqhr}. A possibility is the coupling to low energy spin degrees of freedom in the 2D-bulk. In that respect, it is noteworthy that experimental signatures of a spin-unpolarized 2D-bulk were observed in similar devices set to $\nu=4/3$ \cite{clark1989spinpolarfqhe,tiemann2012spinpolarfqhe}, and that low energy spin excitations were evidenced from the fragile spin polarization at $\nu=1$ \cite{plochocka2009spinpolarnu1}. Another possibility is the coupling to localized electronic states in the 2D-bulk by the long range Coulomb interaction. Such states are more abundant in the fractional quantum Hall regimes, where the fractional gap is not much larger than the energy broadening by disorder. It is conceivable that in our sample, the presence of such states is favored by the wide surface gate located along the edge channel and fixed at ground potential (Fig.~1(b)).

Finally, an important outcome of this work is the demonstration of a direct method to investigate heat transport in the fractional quantum Hall regimes. This opens the path to novel experiments studying the intriguing electronic states found in these regimes.

The authors gratefully acknowledge P. Degiovanni, F. Portier, H. Pothier, P. Roche for discussions. This work was supported by the ERC (ERC-2010-StG-20091028, \#259033).

\emph{Note added.-} Recently, we became aware of two related experimental works investigating heat transport in the quantum Hall regime with quantum dots \cite{venkatachalam2012ltnmqhe,gurman2012encunmfqhr}.


%

\end{document}


\title{Supplementary Material for \\`Chargeless heat transport in the fractional quantum Hall regime'}

\author{C. Altimiras}
\thanks{Current address: CEA, Service de Physique de l'\'Etat Condens\'e (SPEC), 91191 Gif-sur-Yvette, France}
\affiliation{CNRS / Univ Paris Diderot (Sorbonne Paris Cit\'e), Laboratoire de Photonique et de Nanostructures
(LPN), route de Nozay, 91460 Marcoussis, France}
\author{H. le Sueur}
\thanks{Current address: CNRS, Centre de Spectrom\'etrie Nucléaire et de Spectrom\'etrie de Masse (CSNSM), 91405 Orsay Campus, France}
\affiliation{CNRS / Univ Paris Diderot (Sorbonne Paris Cit\'e), Laboratoire de Photonique et de Nanostructures
(LPN), route de Nozay, 91460 Marcoussis, France}
\author{U. Gennser}
\affiliation{CNRS / Univ Paris Diderot (Sorbonne Paris Cit\'e), Laboratoire de Photonique et de Nanostructures
(LPN), route de Nozay, 91460 Marcoussis, France}
\author{A. Anthore}
\affiliation{CNRS / Univ Paris Diderot (Sorbonne Paris Cit\'e), Laboratoire de Photonique et de Nanostructures
(LPN), route de Nozay, 91460 Marcoussis, France}
\author{A. Cavanna}
\affiliation{CNRS / Univ Paris Diderot (Sorbonne Paris Cit\'e), Laboratoire de Photonique et de Nanostructures
(LPN), route de Nozay, 91460 Marcoussis, France}
\author{D. Mailly}
\affiliation{CNRS / Univ Paris Diderot (Sorbonne Paris Cit\'e), Laboratoire de Photonique et de Nanostructures
(LPN), route de Nozay, 91460 Marcoussis, France}
\author{F. Pierre}
\email[Corresponding author: ]{frederic.pierre@lpn.cnrs.fr}
\affiliation{CNRS / Univ Paris Diderot (Sorbonne Paris Cit\'e), Laboratoire de Photonique et de Nanostructures
(LPN), route de Nozay, 91460 Marcoussis, France}

\maketitle

\section{Characterization of the $\nu=4/3$ fractional quantum Hall regime}

Supplementary figure 1 shows the transverse (Hall) and longitudinal resistance of the sample in the vicinity of the $\nu=4/3$ plateau, measured at base temperature $T\simeq40~$mK.

\begin{figure}[tbh]
\centering\includegraphics[width=1\columnwidth]{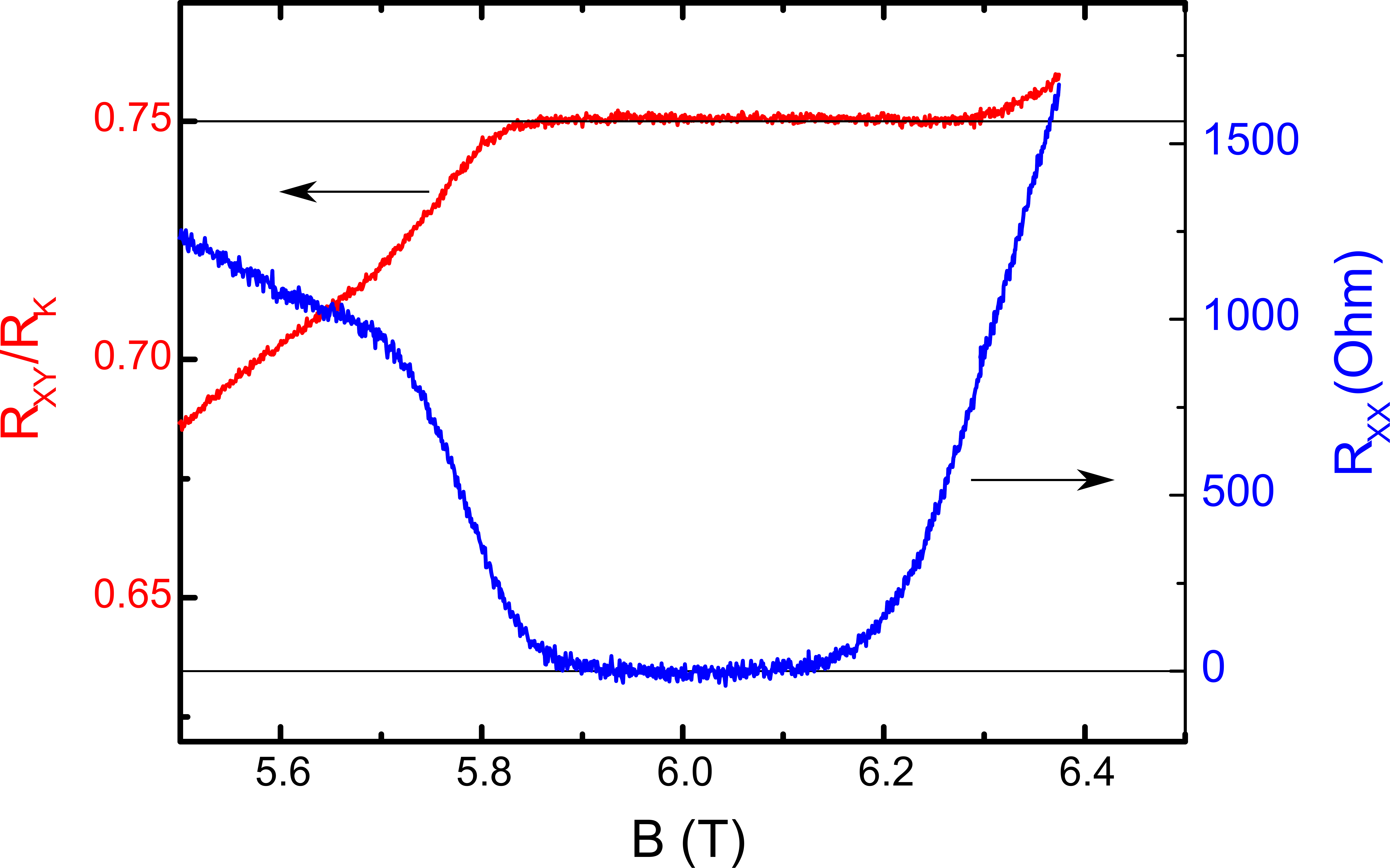}
\caption{Longitudinal ($R_{XX}$) and transverse ($R_{XY}$) resistances plotted as function of magnetic field in the vicinity of the $\nu=4/3$ plateau at $T=40~$mK.}
\end{figure}

\begin{figure}[tbh]
\centering\includegraphics[width=0.85\columnwidth]{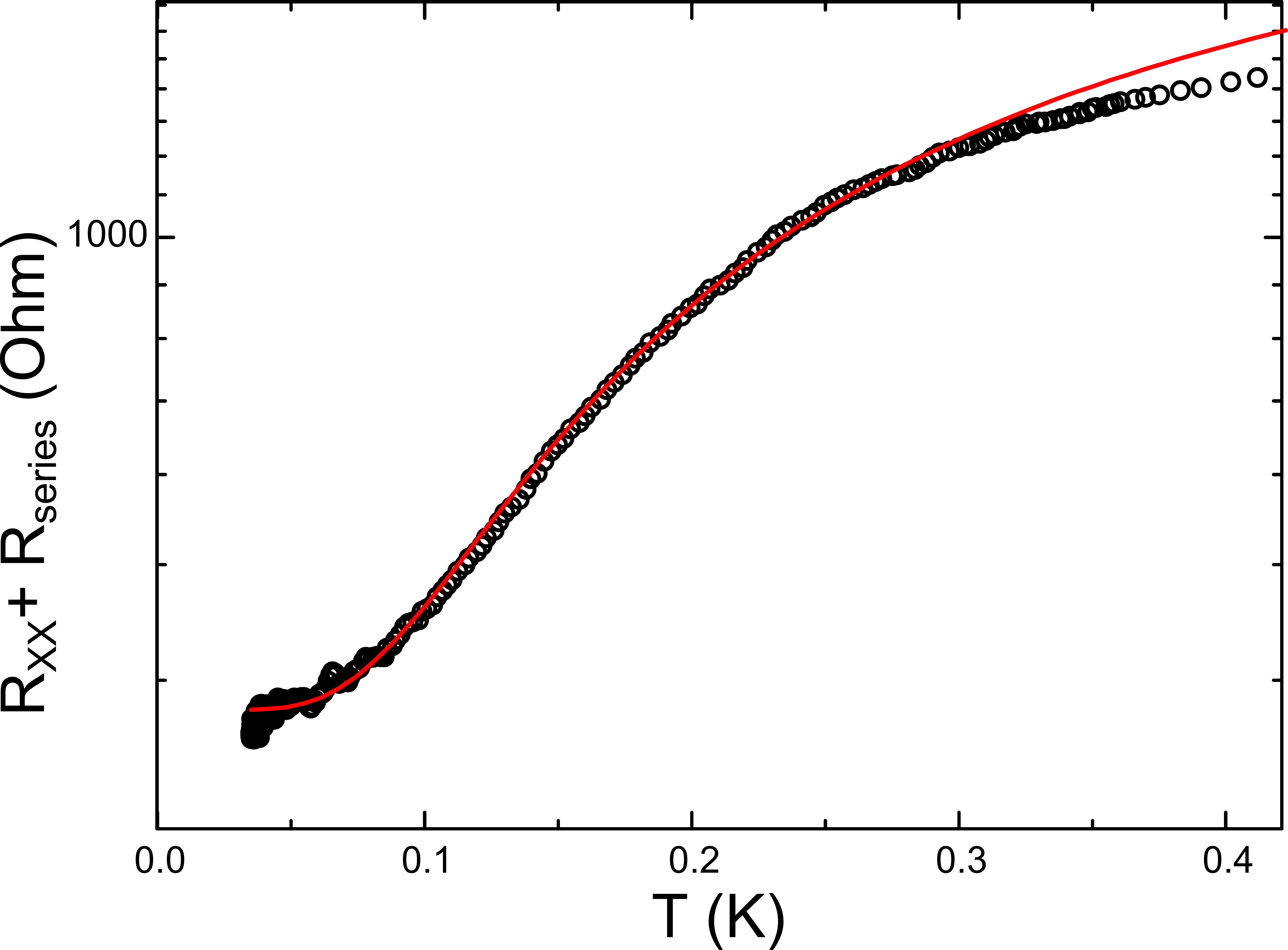}
\caption{Temperature dependence of the longitudinal resistance. The longitudinal resistance $R_{XX}$ in series with a fixed access resistance of $\sim 310~\Omega$ is plotted as symbols in Log scale versus temperature. The continuous line is a fit of the longitudinal resistance with the function $R_0 \exp(\Delta_{4/3}^{eff}/2k_BT)$, using $\Delta_{4/3}^{eff}\simeq k_B \times 700~$mK, and assuming a series resistance of $310~\Omega$.}
\end{figure}

Supplementary figure 2 displays as symbols the longitudinal resistance at the working point $B=6~$T as a function of temperature. The fit of these data with the usual exponential function $R_0 \exp(\Delta_{4/3}^{eff}/2k_BT)$ (continuous line) gives an effective fractional gap $\Delta_{4/3}^{eff}\simeq k_B\times700$~mK. Note that the intrinsic fractional gap $\Delta_{4/3}$ is larger since $\Delta_{4/3}=\Delta_{4/3}^{eff}+\delta E$, with $\delta E$ the energy broadening due to disorder. In our sample we find $\delta E\sim 2.5~\textrm{K}>\Delta_{4/3}^{eff}$ from the onset magnetic field $B\sim 0.2~$T for the Shubnikov de Haas oscillations.

\section{Power injection at the quantum point contact heaters}

\subsection{Derivation of the power injected locally}

\begin{figure}[tbh]
\centering\includegraphics[width=1\columnwidth]{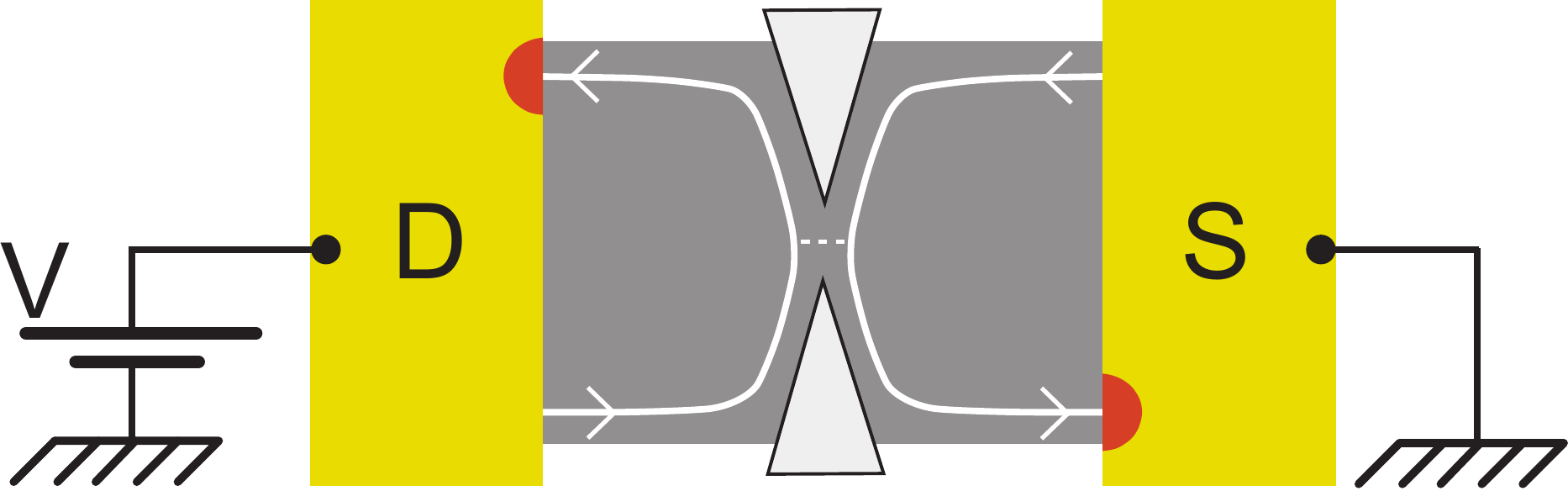}
\caption{Simplified schematic circuit used for power balance considerations. The Landau level filling factor is here set to 1/3. The edge channel is shown as a white line and the propagation direction of the electrical current is indicated by an arrow. Red areas in the reservoirs highlight the locations where the power $P_{\delta\mu}$, associated to the difference in electrochemical potential between edge and reservoir, is dissipated.}
\end{figure}

We derive the power injected locally, at the voltage biased quantum point contact heaters. The analysis does not rely on a detailed description of the fractional edge physics but on general power balance considerations together with the observed chirality of the electrical current (following the analysis detailed in the supplementary information of \cite{altimiras2010nesiqhr} for an integer quantum channel).

We consider the simplified circuit at filling factor 1/3 (one fractional $\nu=1/3$ edge channel) shown in Supplementary figure~3.

The total power provided by the voltage generator is $P=V^2 /(3R_K/\tau)$, with $\tau$ the quantum point contact transmission. This power can be decomposed into two contributions:
\begin{equation}
P=V^2 \tau /3R_K=P_{\delta\mu}+P_{heat}. \label{P}
\end{equation}
The first one ($P_{\delta\mu}$) corresponds to the power injected into the drain and source electrodes due to the electrochemical potential difference $\delta\mu$ with the corresponding incoming edge. The edge electrochemical potential is defined as that of a floating electrode inserted in its path, in the spirit of the `measurement reservoir' model (see e.g. \cite{sivan1986mlfftt}). At unity transmission $\tau=1$, this `electrochemical power' is the only contribution to the dissipated power $P_{\delta\mu} (\tau=1)=P=V^2/3R_K$. In general, the electrochemical power injected by each fractional $\nu=1/3$ edge channel in its output electrode is $(\delta\mu)^2/6h$. At arbitrary transmission $\tau$, the electrochemical potential difference at the input of both the source and drain electrodes is $|\delta\mu|=\tau e |V|$ and one finds:
\begin{equation}
P_{\delta\mu}=\tau ^2 V^2/3R_K. \label{Pmu}
\end{equation}
The second contribution ($P_{heat}$) corresponds to the heat power injected locally, absorbed by excited states on both sides of the quantum point contact. At perfect transmission and reflection, this contribution vanishes.
At intermediate transmissions, $P_{heat}$ is obtained from Supplementary equations~1 and 2:
\begin{equation}
P_{heat}=P-P_{\delta\mu}=\tau (1-\tau) \frac{V^2}{3R_K}. \label{Pedge}
\end{equation}
Half of this power is distributed on each side of the quantum point contact. Consequently, at transmission $\tau=0.5$ the power injected into excited states in one side of the quantum point contact reads $P_{heat}/2\simeq0.25V^2/6R_K$.

\subsection{Practical implementation of local power injection}

As an illustration, we detail here the simultaneous power injection at $H_{U}$ and $H_{D2}$ performed to obtain the data shown in Article Fig. 3(b).

In this case, the edge channels are fully reflected at $H_{D1}$. Since the right electrode across $H_{D2}$ is grounded (see Article Fig. 1(b)), the voltage $V_{HD2}$ is applied by shifting the electrochemical potential of the inner edge channel on the left-hand side of the constriction $H_{D2}$. This is done by biasing $H_{U}$ with symmetric voltage sources $V_{HD2}\pm V_{HU}/2$. Note that the quantum detector is maintained at zero voltage for the probed inner edge channel by applying $V_{HD2}$ to the bottom right voltage source in Article Fig. 1(b).

\begin{figure}[!htph]
\centering\includegraphics[width=1\columnwidth]{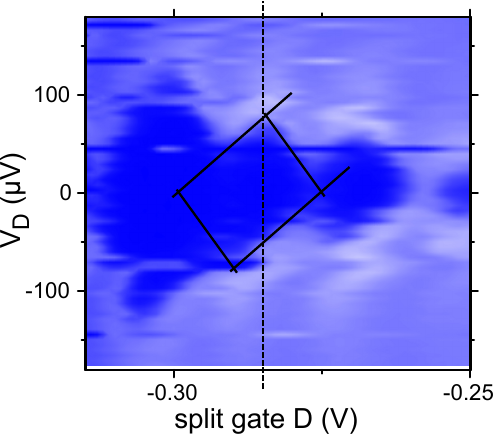}
\caption{Detector dot non-linear characterization. Surface plot of the detector transmission (darkest areas correspond to the transmission $\tau_D=0$). Solid lines delimit the charge stability areas. Data at 40~mK of Article fig.~2(c) are obtained on the dashed line position.
}
\end{figure}

\section{Differences between the present heat detection technique and the energy distribution spectroscopy demonstrated in \cite{altimiras2010nesiqhr}}

We here clarify the differences between the previously demonstrated energy distribution spectroscopy \cite{altimiras2010nesiqhr} and the present heat detection technique.

First, although both the present heat detection technique and that demonstrated in \cite{altimiras2010nesiqhr} make use of a quantum dot, these are tuned in different regimes. In the present work, the quantum dot is tuned in the metallic regime, with no evidences of discrete electronic levels in the dot, whereas in the previous works \cite{altimiras2010nesiqhr,lesueur2010relaxiqhr,altimiras2010ctrlrelaxqhr} the quantum dot is tuned in the single active electronic level regime.

Second, the present heat detection technique relies on activated transport above the Coulomb gap whereas the previous technique \cite{altimiras2010nesiqhr} makes use of the narrow energy filter provided by the single active electronic level in the dot.

In practice, we find that the activated transport technique demonstrated in the present work is well suited to investigate the complex fractional regime, whereas we could not implement on the `fractional' inner edge channel the electronic energy distribution spectroscopy demonstrated in \cite{altimiras2010nesiqhr} due to the stringent constraints on the quantum dot detector (e.g. single electronic level in the quantum dot in an adequate energy window and energy independent quantum dot-edge channel tunnel transmissions).

\section{Quantum dot calibration}

The Coulomb blockade regime of the detector is evidenced from the Coulomb diamond shape differential transmission $\tau_D$ of the `$\nu=1/3$' edge channel versus the detector split gate voltage and applied detector bias voltage $V_D$. The diamonds corresponding to the detector used for the data of Article Fig.~2(c),(d) are shown in Supplementary Fig.~4. The data shown in Article Fig.~2(c) at $T=40$~mK are obtained for the split gate D set to the voltage $-0.284$~V (dashed line).

\begin{figure}[!htph]
\centering\includegraphics[width=1\columnwidth]{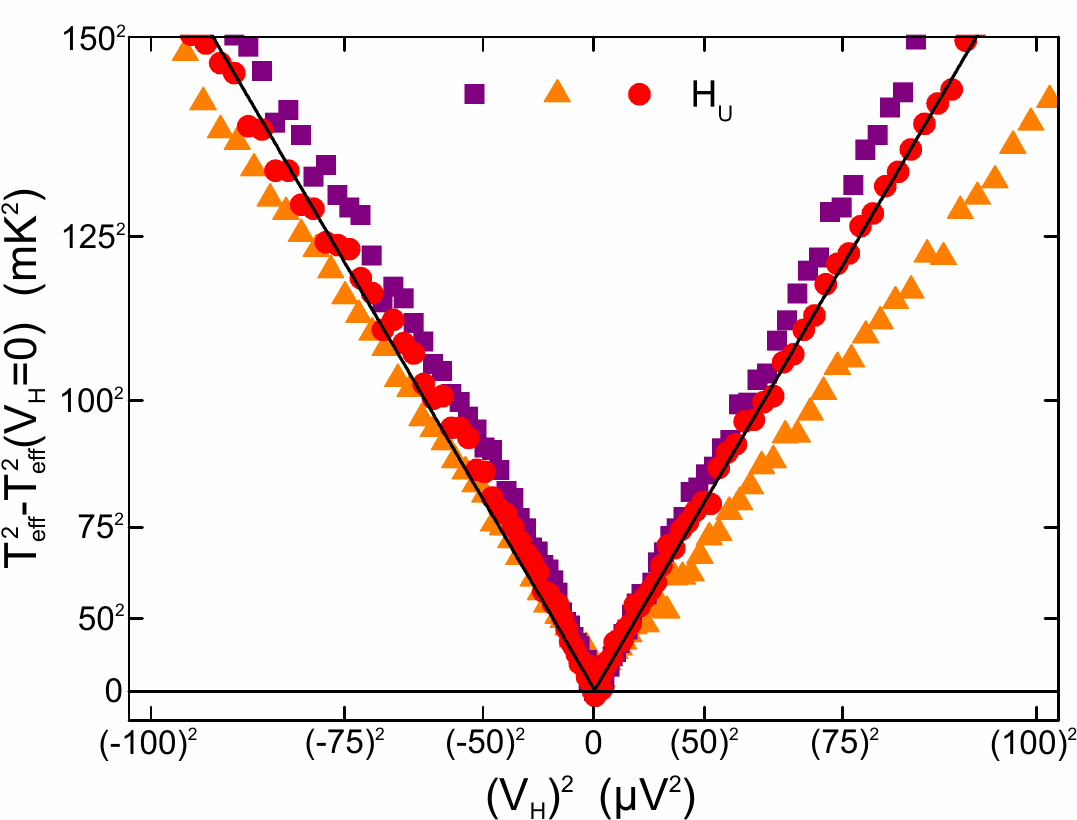}
\caption{Relative precision of heat injection. Difference between the squared effective temperature and the thermal contribution versus $V_H^2$. The red dots correspond to the $H_{D2}$ long edge effect measurement (same data as in Article figure 3(d)), the purple squares to the $H_{D1}$ effect configuration and the orange triangles to the $H_{D2}$ configuration.}
\end{figure}

The coupling of the dot to the drain (source) is characterized by a capacitance $C_D$ ($C_S$) and a resistance $R_D$ ($R_S$). The gate is only capacitively coupled to the dot with $C_G$. From the slopes of the degeneracy lines that delimit the charge stability areas, we estimate $C_D=1.1$ ($C_S=1.4$)~fF and $C_G=1$~aF. These values correspond to a charging energy $E_C=32$~$\mu$eV, consistent with the measured gap displayed Article Fig.~2(c).

Note the absence of lines parallel to the degeneracy lines. This indicates that the detector dot contained a large number of levels separated by $\delta E << k_B T$. Consequently, we use the standard `metallic' dot description of coulomb blockade.

In the regime $\delta E << k_B T < E_C$, the detector calibration at $V_D=0$ is made by changing the base temperature $T$ and measuring the activated differential transmission $\tau_D(T)$. The measured data at the chosen split gate D value are consistent with the standard expression \cite{beenakker1991tcbocqd,kouwenhoven1997etqd}. In the middle of a nearly symmetric metallic dot, this expression reads:
$$\tau_D (T)=\left(\frac{R_K}{R_D+R_S}\right)\frac{1}{\cosh^2(E_C/(2.5 k_B T))}.$$
The activation temperature quoted in the paper is therefore related to the charging energy by $T_C=E_C/(2.5 k_B).$
Note that the above expression is valid assuming that source and drain electrodes and the metallic dot are at thermal equilibrium at temperature $T$ and composed of Fermi quasiparticles. Remarkably, we find an excellent agreement while probing the fractional `$\nu=1/3$' inner channel.

\section{Relative accuracy on the extracted effective temperature $T_{eff}$}

\begin{figure}[!htph]
\centering\includegraphics[width=1\columnwidth]{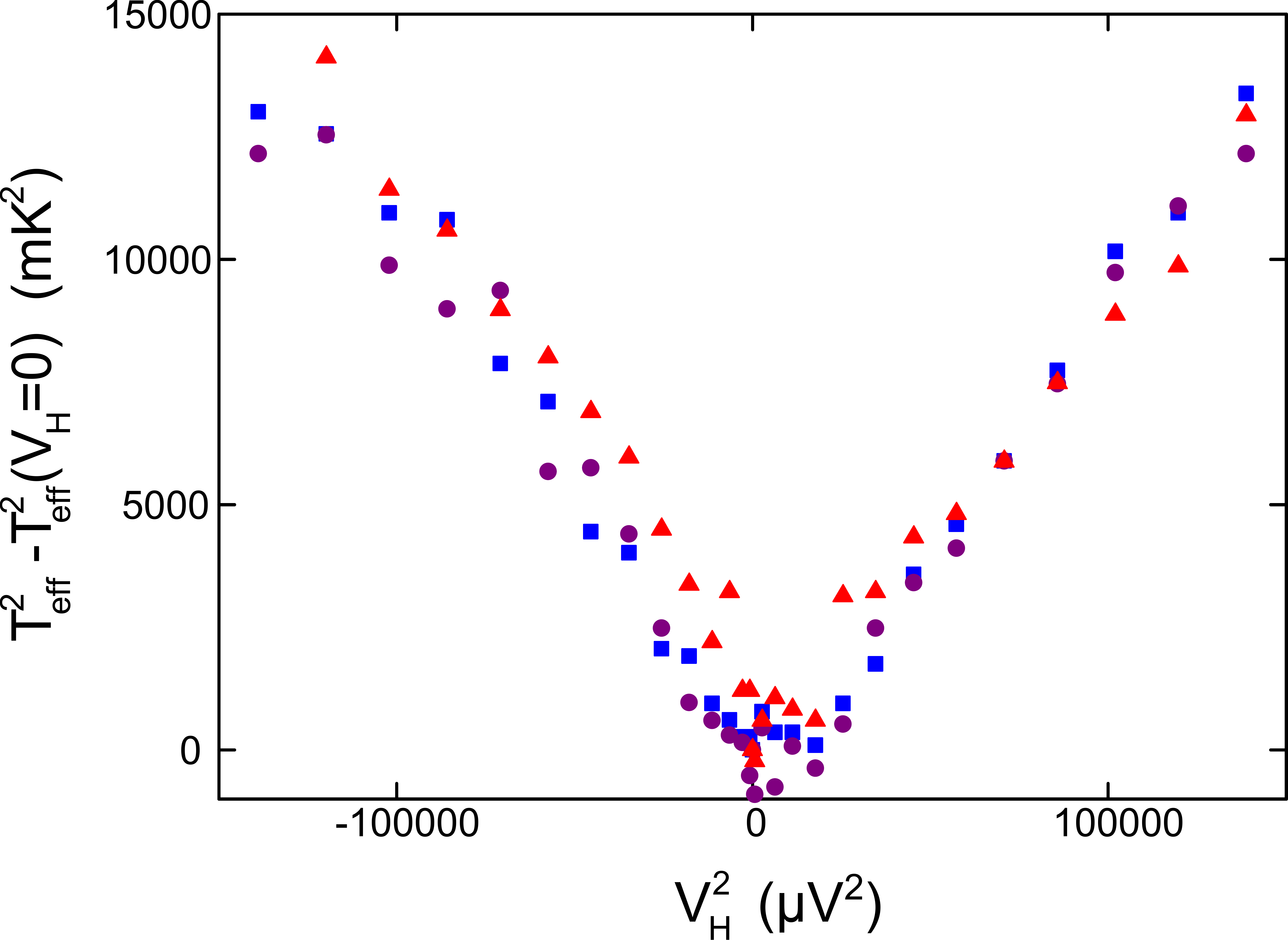}
\caption{Excess squared temperature for different base temperatures versus injected power across the downstream heater $H_{D1}$. Square symbols (blue), circle symbols (purple) and triangular symbols (red) respectively correspond to $T_{base}=50,~100$ and $150$~mK.}
\end{figure}

In order to change the downstream heater used ($H_{D1}$, $H_{D2}$ and $H_{D2}$ long edge), we also change the voltage applied to the corresponding split gates. Due to capacitive cross-talk, this could slightly modify the detector calibration and therefore reduce our relative accuracy when comparing different downstream heaters. However, this can be checked by comparing the heating signals resulting from the same upstream power injection in the different downstream heater configurations: if the quantum dot detector is not modified, the different $H_U$ data should fall on top of each other.

Supplementary figure 5 shows the measured excess square temperature for the three configurations as a function of $V_H^2$. The relative imprecision between the three downstream configurations remains mostly below $15~\%$, which is the relative accuracy quoted in the article.
\\

\section{Experimental test of the procedure used to subtract the thermal background contribution}

We here establish experimentally the procedure to subtract the thermal background.

For this purpose, we compare data taken at strongly different base temperatures, $T_{base}=50,~100$ and $150$~mK. The observation that the resulting excess square temperature $T_{eff}^2(V_H)-T_{eff}^2(V_H=0)$ is essentially independent of the base temperature, at our experimental accuracy of $15 \%$, validates the procedure.

Supplementary figure~6 shows a comparison of the excess square temperature extracted from the detector signal when injecting power at the downstream heater $H_{D1}$ for the three base temperatures. Note that the present test was made on another sample with a different detector (of which diamonds are shown in Article figure~2(b)) but with the same geometry. Similar results are obtained when injecting power upstream on the same sample (data not shown), as well as on a less systematic study but on the same sample and with detector settings as that used in Article figures 2(c),(d) and 3 (data not shown).

\section{How does the present observation of chargeless heat transport away from the edge compare with existing data?}

\subsection{Comparison to neutral edge modes investigations by noise measurements}

Recently, the predicted neutral edge modes were investigated through noise measurements \cite{bid2010obsnm,dolev2011cnmfssll,gross2011unmfquehwcd}. In apparent contradiction with the present work, these measurements have shown no indications of chargeless heat transport through the 2D-bulk.

However, the previous noise experiments \cite{bid2010obsnm,dolev2011cnmfssll,gross2011unmfquehwcd} are not designed to discriminate between chargeless heat transport along the edge and through the bulk. We believe the presence of chargeless heat transport through the 2D-bulk would not have necessarily resulted in a distinct detected noise signal.

Moreover, the distance between power injection and heat detection is larger than in the present experiment by an order of magnitude or more. We found in the present work that increasing the heater-detector distance reduces the upstream heat signal due to chargeless heat transport through the bulk. The much larger distance in the previous noise investigations is likely to strongly reduce the heat signal due to the presently observed chargeless heat transport through the bulk.

Other possibly important differences include the different heat detection schemes (noise measurement vs thermally activated current) and the power injection method (in Refs.~\cite{bid2010obsnm,dolev2011cnmfssll,gross2011unmfquehwcd} the power injection to neutral modes is performed with a current biased ohmic contact, possibly due to the hot spot associated with an incoming charge mode).

We therefore conclude that the present observation of a chargeless heat transport through the bulk is not incompatible with the previous experiments~\cite{bid2010obsnm,dolev2011cnmfssll,gross2011unmfquehwcd}.

\subsection{Heat transport investigations at integer filling factors}

\subsubsection{Filling factor $\nu=1$}

In the experiment performed by Granger and coworkers \cite{granger2009och}, the heat current was found to obey the same chirality as the electrical current, for relatively long distances between power injection and heat detection ($20-60~\mu$m). In addition, an apparent energy leakage was observed, whose mechanism remains unresolved. The presently observed chargeless heat transport through the bulk could be related to the apparent energy leakage in \cite{granger2009och}, possibly due to similar low energy magnetic excitations in the 2D-bulk \cite{plochocka2009spinpolarnu1}.

\subsubsection{Filling factor $\nu=2$}

No similar chargeless heat transport was detected on the same sample at filling factor $\nu=2$ \cite{altimiras2010nesiqhr,lesueur2010relaxiqhr,altimiras2010ctrlrelaxqhr}.

It is directly seen in the experiment shown Fig.~2(b) of \cite{altimiras2010ctrlrelaxqhr}, which is the equivalent, but in the `forward' direction, of the measurement labeled `$H_{D2}$ long edge' in Article Fig.~3(d). In the experiment at $\nu=2$, one finds no discernable heat transfer across the same constriction $H_{D1}$ of the same sample when the electrical path is deviated toward a cold reservoir ($G_i=1$ in Fig.~2(b) of \cite{altimiras2010ctrlrelaxqhr}).

The conclusion that no significant amount of heat was deviated toward extra modes at the power injection point and on sub-micron length scales can also be reached from the fact that the injected power is fully recovered downstream the power injection point at our experimental accuracy (see Fig.~4d in \cite{altimiras2010nesiqhr}).

\section{Supplementary discussions on heat transport mechanisms}

\subsection{Coupling to phonons}

In principle it is conceivable that electron-phonon coupling is responsible for the detected chargeless heat transport through the bulk. However we believe it is unlikely, as detailed below.

Our arguments rely on theoretical considerations, on the actual observation that this mechanism is negligible at filling factor $\nu=2$ for similar length and energy scales, and on the relatively small heating detected on the `cold' side of a quantum dot detector.

\subsubsection{Theoretical expectations}

The expected exponential decay of the electron-phonon coupling along quantum Hall channels and the small power transfer from electrons to phonons in 2D electron gases seem incompatible with an explanation relying on electron-phonon interactions for the chargeless heat transport observed in the present work.

\textbf{Temperature dependence}

At low temperature along a quantum Hall edge channel, the electron-phonon coupling is expected to be exponentially suppressed with a characteristic inelastic time proportional to $\exp(-T_C/T)$, and a cross-over temperature evaluated to $T_C\approx 4~$K for our sample \cite{martin1990ephqhr}.

This prediction seems incompatible with the observation that heating upstream scales with the power injected downstream (see Article Fig. 3(d)).

\textbf{Quantitative estimate in 2D electron gases at low temperature}

We are not aware of quantitative theoretical predictions for the electron-phonon coupling along quantum Hall channels. In order to get an order of magnitude estimate of the power transferred from electrons to phonons, we will use the theoretical predictions for a 2D electron gas in the ballistic regime.

According to theory \cite{price1982eph2deg} and experiment \cite{mittal1996eph2deg}, in Ga(Al)As ballistic 2D electron gases at sub-Kelvin temperatures the power transferred from electrons to phonons reads:
\begin{equation}
P_{e-ph}=1.65\times10^6n_S^{-1/2}A(T_e^5-T_{ph}^5),
\end{equation}
with $n_S$ the electron density per unit area ($n_S\simeq2\times10^{15}~\mathrm{m}^{-2}$), $A$ is the area, $T_e$ is the electronic temperature and $T_{ph}$ is the phonon temperature.

In order to have a significant electron-phonon contribution to the heating detected upstream, the power transferred to phonons should be at least $1/10$ of the injected power (note that, in addition, the phonons should be significantly heated up in order to transmit back to the electronic system some of the absorbed power). An upper bound of the power transferred from electrons to phonons is obtained using the above expression with the highest effective electronic temperature $200~$mK measured downstream for $V_{HU}=100~\mu$V, a cold phonon bath at $50~$mK and assuming a large edge channel of width $1~\mu$m. Under these conditions, a macroscopic length of $0.14~$mm is required to transfer $1/10$ of the injected power from electrons to phonons. This is in contrast with the small heater-detector edge distances, below 3 microns in the present experiment.

\subsubsection{Electron-phonon coupling at filling factor $\nu=2$}

Several observations at filling factor $\nu=2$ on the same sample show that electron-phonon energy exchanges are small (these conclusions also apply to the coupling with the low energy electronic states in the surface metal gates).

The most direct demonstration is the observation in \cite{altimiras2010ctrlrelaxqhr} that all inelastic mechanisms along a quantum edge channel can be frozen over a propagation distance of $8~\mu$m, for an effective electronic temperature of 85~mK. In particular, this observation shows that the energy exchanges between electrons and phonons have a negligible effect on the non-equilibrium electronic energy distribution function for this propagation length and energy scale.

Moreover, we find that an edge channel driven out-of-equilibrium relaxes toward the same hot Fermi function (of electronic temperature up to 110~mK on top of a bath temperature of 40~mK) after a propagation distance of either $10~\mu$m or $30~\mu$m \cite{lesueur2010relaxiqhr}. The fact that this hot Fermi function remains at the same temperature over at least $20~\mu$m shows that no significant net power is transferred toward the phonons degrees of freedom. Assuming the phonons are at cold equilibrium, this observation implies that energy exchanges between electrons and phonons are negligible on these length and energy scales.

\subsubsection{Electron-phonon coupling at filling factor $\nu=4/3$}

An indication that electron-phonon interactions do not result in significant heat transfers in the present experiment at $\nu=4/3$ is the supplementary data set showing that, at our experimental accuracy, only one side of the quantum dot detector is heated up. See the dedicated subsection below.

\subsection{Heat paths discussion}

\subsubsection{Isotropic chargeless heat transport through the bulk?}

We demonstrate experimentally the presence chargeless heat transport located further in the bulk than the electrical edge path. One may ask if this heat current flows in all directions within the 2D-bulk.

Note first that far enough inside the 2D-bulk the edge has no influence and therefore there can be no preferred direction relative to the electrical current along the edge.

In addition, the observation of a similar upstream heat signal when the power injected on $H_{D1}$ and $H_{D2}$ is scaled as the heater-detector distance seems consistent with an isotropic 2D-bulk heat transport. Indeed, in the stationary regime and ignoring interactions with other degrees of freedom, the power injected locally in the bulk is equal to the outgoing energy current across a perimeter enclosing the power injection point. For an isotropic chargeless heat transport in the 2D-bulk, the corresponding heat current is distributed equally at a given distance and therefore scales with the injected power and inversely with the distance to power injection. Note that such scaling is only approximately valid for the geometry of the studied sample. At the investigated heater-detector distances using $H_{D1}$ and $H_{D2}$, the injected power would redistribute roughly on a quarter of circle.

\subsubsection{Chiral heat transport along the edge?}

One may ask if heat transport occurs only in the bulk, or also along the edge in the `forward' direction (with the same chirality as the electrical current). Our findings very strongly suggest that heat transport in the `forward' direction is also carried along the edge and that it is the main heat transport mechanism in that direction.

First, the very observation of chiral charge transport along the edge points out a forward heat current along the edge. Indeed, to the best of our knowledge, according to theory a chiral charge transport along the quantum Hall edge is always associated with the propagation of electronic excitations along the edge that transfer heat in the same direction as the electrical current (see e.g. \cite{wen1992tesfqh,kane1997qttfqhe,chamon1994er}). (In addition, other excited states could transport heat along different paths, possibly through the bulk due to e.g. Coulomb interaction or spin polarization.)
Second, we find that heat transport is much more efficient in the forward direction than in the backward direction. (We can compare the detected heating signal for heat injection points at similar edge distances upstream and downstream. We show in Article Fig. 3(d) that it is necessary to inject about four times more power when using the downstream heat injection point $H_{D1}$ in order to obtain the same heating signal as using $H_U$). This suggests the presence of an extra heat transport mechanism in the forward direction, in addition to the detected chargeless heat transport through the bulk observed directly in the backward direction. This therefore corroborates the expected forward heat transport along the edge.

\section{Supplementary heat transport data at $\nu=4/3$}

In this section we present supplementary data providing information on the chargeless heat transport mechanism revealed in the present work.

\subsection{Heating of the `fractional' inner edge channel with power injected in the `integer' outer edge channel}

We here show data demonstrating that the inner edge channel can be heated up by injecting power in the outer edge channel. Remarkably, a similar heating is observed for upstream or downstream power injection at, respectively, $H_{U}$ and $H_{D1}$ which are located approximately at the same distance from the detector $D$. This suggests that the main heat transfer mechanism responsible for the presently observed heating is not a direct coupling between inner and outer edge channels, but rather the coupling of both channels with other low energy modes.

\begin{figure}[!htph]
\centering\includegraphics[width=1\columnwidth]{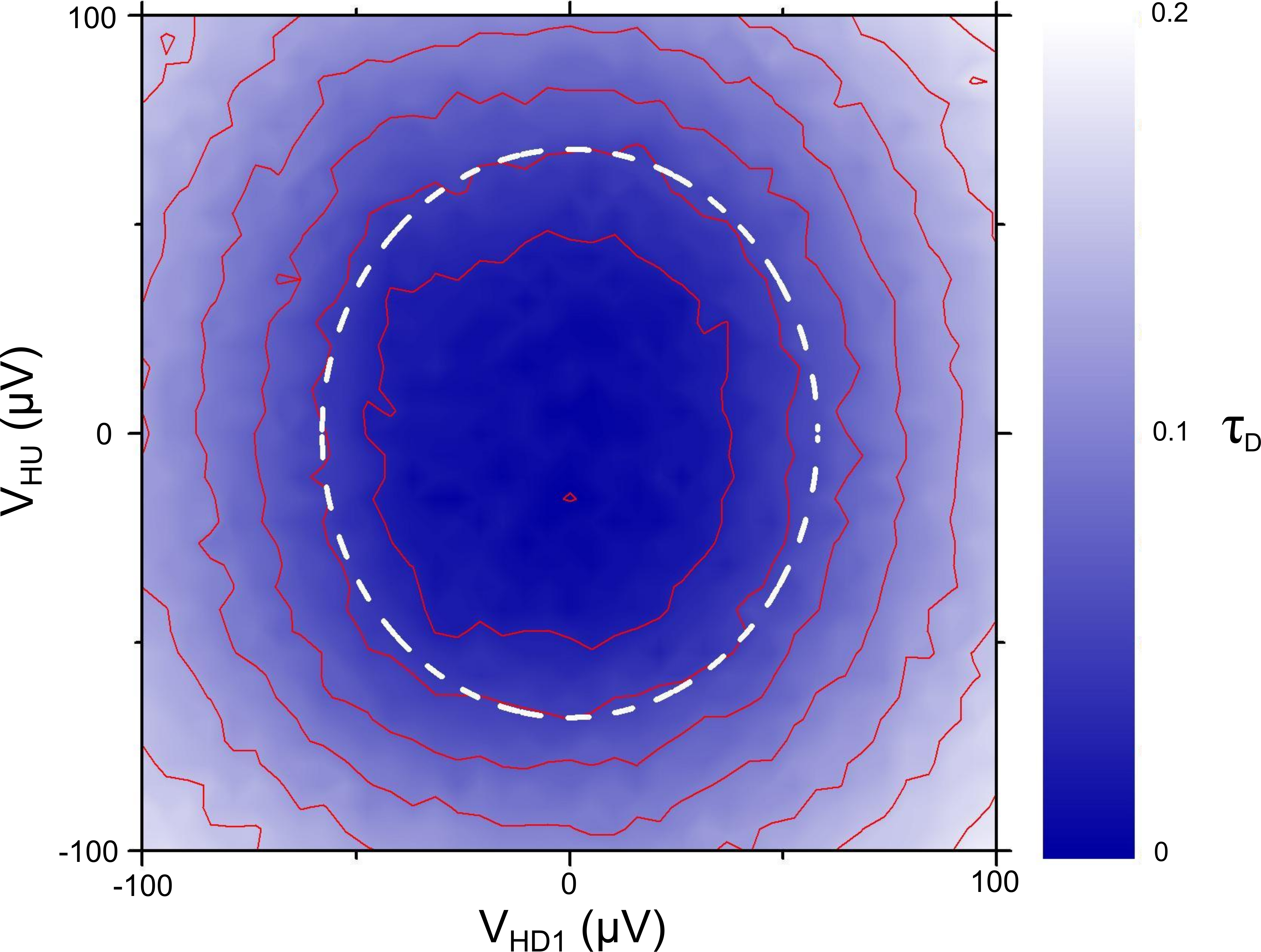}
\caption{Inner edge channel heat signal versus outer edge channel power injection. The detector inner edge channel transmission is plotted versus the simultaneously applied upstream $V_{HU}$ and downstream $V_{HD1}$ heater voltages. Continuous lines (red) are equitransmission contours at integer multiples of $0.025$. The white dashed line is a fit of the equitransmission line $\tau_D=0.05$ with an ellipse of minor (major) diameter $58~\mu$V ($68~\mu$V) along $V_{HD1}$ ($V_{HU}$).}
\end{figure}

Supplementary Fig.~7 shows the activated transmission probability of the `fractional' inner edge channel across the quantum dot detector $D$, plotted as a function of the applied voltage across the heaters $H_{U}$ and $H_{D1}$ set to inject power in the `integer' outer edge channel. For this purpose, in contrast with the data displayed in the article, the constrictions $H_{U}$ and $H_{D1}$ are here tuned to transmit half of the outer edge channel (with the inner edge channel fully reflected). Note also that the detector is set to a different operating point that in the article, here with a blockaded transport up to a onset drain-source voltage of $35~\mu$V at $30~$mK (similar data were obtained at another operating point with a half smaller onset drain-source voltage).

The increased detector transmission probability with power injection at $H_{U}$ or/and $H_{D1}$ demonstrates that heat can be transferred from the outer edge channel to the inner one. Remarkably, in contrast with observations when heat injection and detection are performed on the same inner channel (see Article Fig.~3(d)), the same heating (increase in detector transmission) is here obtained with a slightly lower power injection from the downstream heater $H_{D1}$ than from the upstream heater $H_U$. This is illustrated by comparing the equipotential line $\tau_D=0.05$ with the ellipse $(V_{HU}/68~\mu \textrm{V})^2+(V_{HD1}/58~\mu \textrm{V})^2=1$ (white dashed line). Interestingly, the downstream heater $H_{D1}$ is at an edge distance of $1.4~\mu$m from the detector, slightly closer than $H_U$ located at an edge distance of $1.8~\mu$m. Assuming that more heat propagates downstream along the `integer' outer edge channel and taking into account the other observation that power injected in the inner channel propagates preferentially downstream, this finding appears incompatible with a direct heat transfer between outer and inner channel along the edge. Instead, this suggests that the inner and outer edge channel are both coupled to some other low energy modes that do not contribute to charge transport and that propagate heat equally upstream and downstream.

\subsection{Heating in the `integer' outer edge channel probed with a quantum dot detector in the discrete electronic level regime}

We here show the heating signal detected on the outer edge channel when the power is injected upstream at a short distance ($0.7~\mu$m) in the same channel (see Supplementary Fig.~8). Importantly, the quantum dot detector is set in the discrete electronic level regime. This allows us to probe separately the temperatures in the outer edge channels on both sides of the dot. We observe an important heating on the same drain side where power is injected whereas the source side remains cold, within our typical error bars $\pm 10\%$. This observation, which corroborates our findings at $\nu=2$, shows that heat transfers mediated by phonons in the substrate or by electronic degrees of freedom in the surface metal gates are much smaller than the downstream heat transfers detected in the drain outer edge channel. However, the accuracy of the present test is insufficient to directly and unambiguously rule out the role of phonons and surface metal gates regarding the detected chargeless heat transport through the bulk.

\begin{figure}[!htph]
\centering\includegraphics[width=0.85\columnwidth]{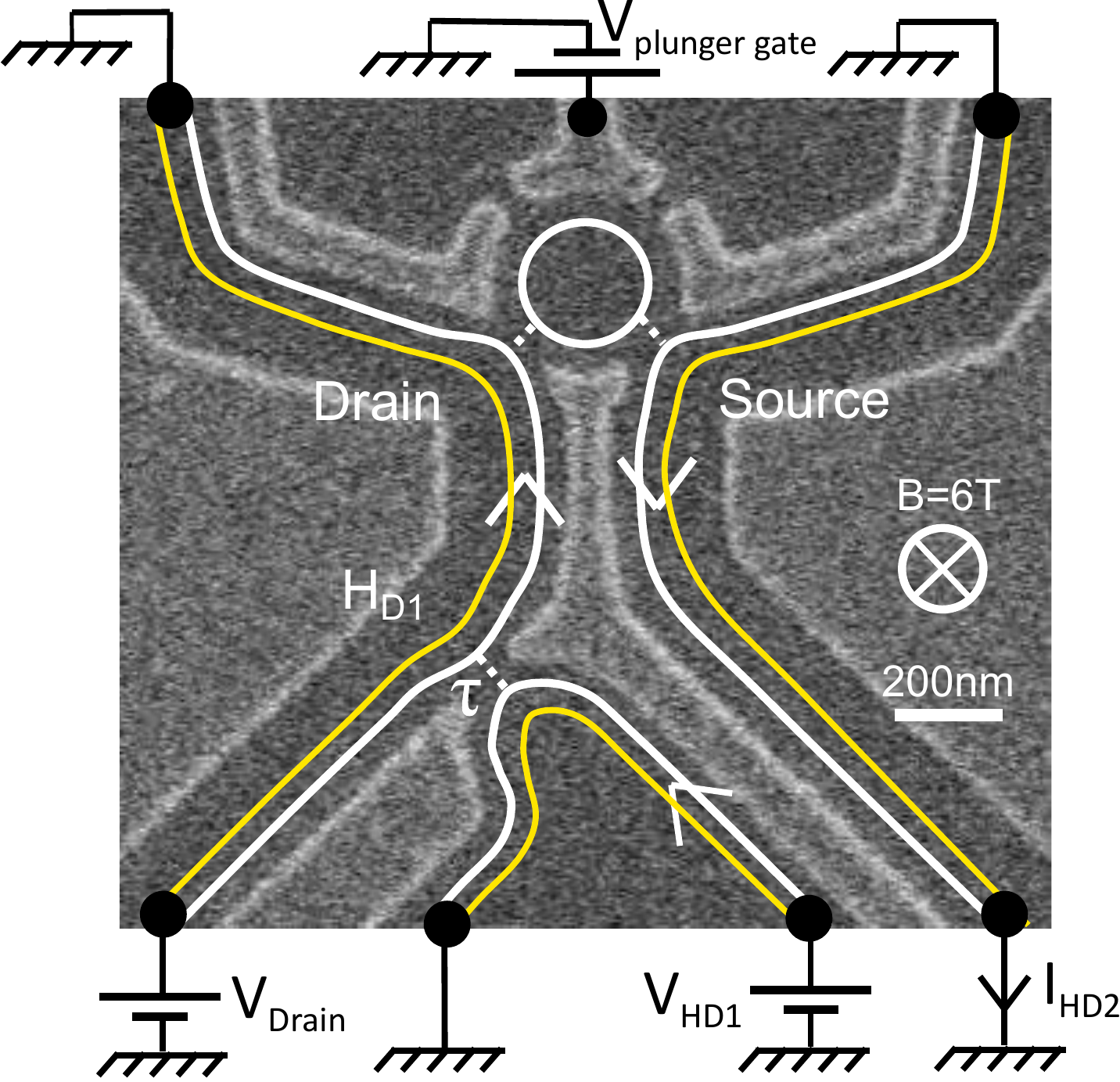}
\caption{Experimental configuration corresponding to the data displayed in Supplementary figure 9.}
\end{figure}

The quantum dot detector is here at the location of $H_{D2}$ and tuned using the four nearby surface metal gates (see Supplementary Fig.~8). The detector is set to probe the `integer' outer edge channel. Power was injected at $H_{D1}$ (drain side of the quantum dot) in the same outer channel. For this purpose the constriction $H_{D1}$ is set to transmit half of the outer channel (the inner channel being fully reflected).

\begin{figure}[!htph]
\centering\includegraphics[width=1\columnwidth]{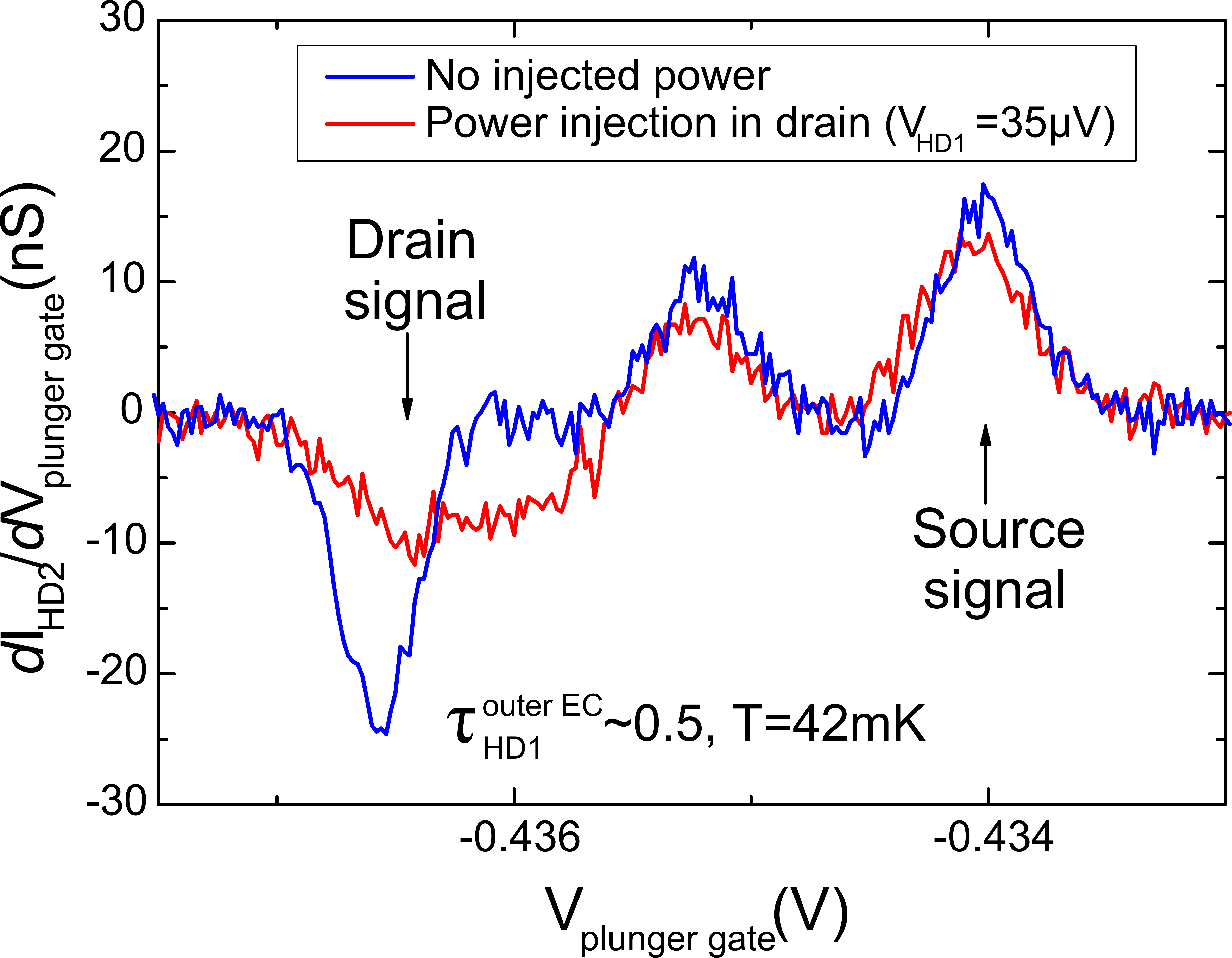}
\caption{Outer edge channel heat detection in source and drain based on a quantum dot detector set in the discrete electronic level regime. The quantum dot detector is here located at $H_{D2}$ and tuned using the nearby top metal gates (see our previous works on the same sample \cite{altimiras2010nesiqhr,lesueur2010relaxiqhr,altimiras2010ctrlrelaxqhr}). Although the detector has more than one active electronic level in the energy window defined by the applied drain-source voltage, the width of the displayed left dip (right peak) probes the temperature of the 'integer' outer edge channel in the drain (source) electrode. When some power is injected in the probed outer edge channel at $H_{D1}$ on the drain side, we observe a rounding of the left dip associated with an effective `temperature' increase at the detection point on the drain side (and a shift due to the related change in electrochemical potential). On the contrary, the right dip associated with the source temperature does not change significantly with injected power.}
\end{figure}

The signal displayed in Supplementary Fig.~9 corresponds to the variation of the current across the quantum dot detector when the plunger gate voltage is changed. As discussed in further details below, the width of the outer-most left dip (right peak) is an increasing function of the temperature in the drain (source) outer edge channel. Note that the extra center peak results from the presence of two active quantum dot levels. The observation that the left dip is strongly widened when a voltage bias $V_{HD1}=35~\mu$V is applied across $H_{D1}$ shows that the `effective' temperature of the drain outer edge channel is increased. In contrast, the right peak aspect ratio is mostly unchanged by power injection in the drain side. Consequently, the temperature of the outer edge channel in the source side of the quantum dot is not significantly heated up by power injection on the other side of the dot.

This finding shows in the present experimental configuration, at $\nu=4/3$, that heat transfers across a narrow area depleted by a voltage biased surface gate are much smaller than heat transfers toward a downstream location along the edge channel where power is injected. However, the accuracy of this test does not allow us to rule out directly and unambiguously the possibility that the phonons or the surface metal gates play a role in the chargeless heat transport observed upstream power injection.

We now provide further detail on heat detection with a discrete electronic level quantum dot. In the simple case of a canonical quantum dot with a single active electronic level in the probed energy window (approximately delimited by the drain-source voltage, here $V_{Drain}=-108~\mu$V), the left dip (right peak) is directly proportional to the derivative of the electronic energy distribution function in the drain (source) and the plunger gate voltage is proportional to the energy \cite{altimiras2010nesiqhr}. With equilibrium Fermi distributions, the half-width of the dip (peak) is then simply proportional to the temperature in the drain (source).
The present configuration is more complex since there are at least two active electronic levels, as evidenced by the presence of two peaks. Nevertheless, we expect that the left dip and right peak probe mostly the outer edge channel of, respectively, the drain and the source. Moreover, we expect that the peak/dip width remains an increasing function of the corresponding temperature, at least until the different peaks/dips merge. Note that simple fits of the peak/dip at cold equilibrium (`No injected power' data in Supplementary Fig.~9) using the derivative of a Fermi function gives the fit temperatures $T_{fit}=51~$mK, $60~$mK and $58~$mK respectively for the left dip, center peak and right peak (`No injected power' data in Supplementary Fig.~9). The same fit of the right peak in presence of power injection in the drain ($V_{HD1}=35~\mu$V) gives $T_{fit}=68~$mK. With a typical temperature error bar of approximately $\pm10\%$ for each temperature, the difference from the corresponding cold equilibrium fit temperature $58~$mK is below our experimental accuracy.


%